\documentclass{aa}
\newcommand{\bit}{\begin{itemize}}
\newcommand{\eit}{\end{itemize}}
\newcommand{\ben}{\begin{enumerate}}
\newcommand{\een}{\end{enumerate}}
\newcommand{\bde}{\begin{description}}
\newcommand{\ede}{\end{description}}

\usepackage{graphicx}
\usepackage{lscape}

\begin{document}
\title{Dark matter density profiles: A comparison of nonextensive theory with N-body simulations}
\author{T. Kronberger$^{1,2}$,
        M. P. Leubner$^{1}$, and
        E. van Kampen$^{1}$
        }

\institute{ $^1$Institut f\"ur Astro- und Teilchenphysik,
            Leopold-Franzens-Universit\"at Innsbruck,
            Technikerstr. 25,
            A-6020 Innsbruck, Austria\\
            $^2$Institut f\"ur Astrophysik,
            Universit\"at G\"ottingen,
            Friedrich-Hund-Platz 1,
            D-37077 G\"ottingen, Germany}

\offprints{T. Kronberger, \email{Thomas.Kronberger@uibk.ac.at}}

\date{-/-}

\abstract{Density profiles of simulated galaxy cluster-sized dark
matter haloes are analysed in the context of a recently introduced
nonextensive theory of dark matter and gas density distributions.
Nonextensive statistics accounts for long-range interactions in
gravitationally coupled systems and is derived from the
fundamental concept of entropy generalisation. The simulated
profiles are determined down to radii of $\approx$ 1\% of
R$_{200}$. The general trend of the relaxed, spherically averaged
profiles is accurately reproduced by the theory. For the main free
parameter $\kappa$, measuring the degree of coupling within the
system, and linked to physical quantities as the heat capacity and
the polytropic index of the self-gravitating ensembles, we find a
value of $-15$. The significant advantage over empirical fitting
functions is provided by the physical content of the nonextensive
approach.

 }
\authorrunning {Kronberger, Leubner \& van Kampen}
\titlerunning {Dark matter density profiles}
\maketitle
%
%________________________________________________________________

\section{Introduction}

The radial density profiles of dark matter (DM) haloes, generated
in the process of hierarchical structure formation, were modelled
in the past primarily on phenomenological grounds.
High-resolution N-body simulations were performed (e.g. Navarro et
al. \cite{Navarro96,Navarro97}; Moore et al. \cite{Moore99}) in
order to reveal the DM density distribution. The functional
dependence
$\rho_{\rm{DM}}\propto(r/r_S)^{-\alpha}(1+r/r_S)^{-(3-\alpha)}$
(Zhao \cite{Zhao96}) provides reasonable fits to haloes on all
observable scales (e.g. Ricotti \cite{Ricotti03}). Here $r_S$ is a
scaling radius and the parameter $\alpha$ is related to the
initial power spectrum. For $\alpha=1$ this expression represents
the Navarro, Frenk \& White (\cite{Navarro96,Navarro97}),
hereafter NFW profile, while the Moore et al. (\cite{Moore99})
profile follows with $\alpha=1.5$. Both on theoretical and on
observational grounds there is still profound discussion on the
innermost slope of the DM density profile. While numerical
simulations generally predict a divergent, 'cuspy' behaviour
towards the centre (e.g. Navarro et al. \cite{Navarro04})
observations favour a flat, cored profile (e.g. Gentile et al.
\cite{Gentile04} and references therein).

Only a few attempts to provide theoretically motivated models for
density profiles of cosmic structure were proposed so far. Early
analytic work was pioneered by Gunn \& Gott (\cite{Gunn72}) and
subsequently elaborated by Fillmore \& Goldreich
(\cite{Fillmore84}) and Bertschinger (\cite{Bertschinger85}), who
derived similarity solutions for the self-similar collapse of
spherical perturbations. Those solutions were naturally power-laws
for virialised objects. Later, correlations between the slope of
the density profile and the form of the power spectrum of the
initial density perturbations were recognized (e.g. Hoffman
\cite{Hoffman88}). More recent attempts based on analytic infall
models were provided by Williams et al. (\cite{Williams04}) and
Ascasibar et al. (\cite{Ascasibar04}) while Hansen
(\cite{Hansen04}) and Austin et al. (\cite{Austin05}) used the
Jeans equation to gain further insight.

In this work we apply a recently proposed nonextensive theory for
DM and gas density profiles (Leubner \cite{Leubner05}),
accounting for long-range interactions and correlations present in
any astrophysical system, to results obtained from N-body
simulations. The term 'nonextensive' is used for statistical
ensembles, whose total entropy is not an additive quantity as in
standard Boltzmann-Gibbs-Shannon (BGS) thermo-statistics. Instead,
pseudo-additive entropy generalisation accounts for correlations
between sub-systems. The possibility to describe DM density
profiles using nonextensive statistics was recently also suggested
(Hansen et al. \cite{Hansen05}) in the context of Eddingtons
formula (e.g. Binney \& Tremaine \cite{Binney94}), whereas an
earlier theoretical link between entropy generalisation and
self-gravitating structures was provided by Plastino \& Plastino
(\cite{Plastino93}).

The classical BGS extensive thermo-statistics constitutes a
powerful tool when microscopic interactions and memory are short
ranged and the environment is an Euclidean space-time, a
continuous and differentiable manifold. However, in the present
situation we are dealing with astrophysical systems, generally
subject to spatial or temporal long-range interactions, i.e.
ensembles evolving under correlations that makes their behavior
nonextensive. A suitable generalization of the BGS entropy for
statistical equilibrium was first proposed by Renyi
(\cite{Renyi55}) and subsequently by Tsallis (\cite{Tsallis88}),
preserving the usual properties of positivity, equiprobability and
irreversibility, but suitably extending the standard extensivity
or additivity to nonextensivity. The main theorems of the
classical Maxwell-Boltzmann statistics admit profound
generalizations within nonextensive statistics (sometimes referred
to as q-statistics where $q$ characterizes the degree of
nonextensivity of the system), wherefore a variety of subsequent
analyses were devoted to clarify the mathematical and physical
consequences of pseudo-additivity, for an early review see e.g.
Tsallis (\cite{Tsallis95}). Those include a reformulation of the
classical N-body problem within the extended statistical mechanics
(Plastino et al. \cite{Plastino94}) and the development of
nonextensive distributions (Silva et al. \cite{Silva98}; Almeida
\cite{Almeida01}). A deterministic connection between the
generalized entropy and the resulting power-law functionals was
recognized (Andrade et al. \cite{Andrade02}), as well as the dual
structure of nonextensive statistical theory (Karlin et al.
\cite{Karlin02}).

Besides new insights into fundamental physics astrophysical applications
support q-non-extensive statistics, proposing distribution functions for
stellar polytropes (Plastino et al. \cite{Plastino93}) and explaining the solar neutrino
counting rate (Kaniadakis et al. \cite{Kaniadakis96}). Cosmological
tests provide a bound on the degree of possible non-extensivity from
primordial helium abundance data (Torres et al. \cite{Torres97}) and the temperature of
the cosmic microwave background radiation in a Robertson-Walker universe
was shown to be independent of the degree of non-extensivity (Hamity et al. \cite{Hamity96}).
Furthermore, generalized statistics was applied to large scale
astrophysical systems subject to long-range gravitational forces in
view of galaxy distribution studies (Nakamichi et al. \cite{Nakamichi02}) and reformulated
also in the context of special relativity (Kaniadakis \cite{Kaniadakis02}). Recently
nonextensive theory was successfully introduced also to study the scale dependence
of intermittent flows in astrophysical plasma turbulence, appearing as
consequence of long-range interactions (Leubner and V\"{o}r\"{o}s \cite{LeubnerV05}).
In this context the resulting bi-kappa distribution function appears as manifestation
of the dual nature of nonextensive statistics, which provides also the physical
background of entropy bifurcation in the theoretical context of DM and gas
density distributions of clustered matter (Leubner \cite{Leubner05}).

We relate in the following nonlocality in gravitationally
clustered astrophysical structures to the presence of long-range
forces in nonextensive systems and demonstrate that density
distributions derived within the framework of entropy
generalisation consistently reproduce the density profile of
cluster-sized dark matter haloes forming in $\Lambda$ cold dark
matter ($\Lambda$CDM) N-body simulations. This result is
furthermore supported by the integrated mass profile of an
observed relaxed galaxy cluster. Consequently, nonextensive
statistics provides a physically interpretable alternative to
empirical fitting procedures.

\section{Theory}

Let us first illuminate the property of pseudo-additivity in the
context of nonextensive entropy generalization by considering two
sub-systems $A$ and $B$ such that

\begin{equation}
S_{\kappa}(A+B)=S_{\kappa}(A)+S_{\kappa}(B)+\frac{1}{\kappa} S_{\kappa}(A)S_{\kappa}(B)
\label{1}
\end{equation}

where $S_{\kappa}$ denotes the entropy as depending on the
entropic index $\kappa$, a parameter specifying the degree of
nonextensivity in the system. For $\kappa = \infty$ the last term
on the right hand side cancels leaving the additive entropy of the
standard BGS statistics. Hence, nonlocality or long-range
interactions are introduced by the multiplicative last term on the
right hand side of Eq. (\ref{1}) accounting for correlations
between the subsystems $A$ and $B$. As measure of the entropy
mixing the parameter $\kappa$ quantifies the degree of statistical
correlations in the system and thus accounts for nonlocality and
long-range interactions or couplings, respectively. In general,
the pseudo-additive, $\kappa$-weighted term may assume positive or
negative definite values indicating a nonextensive entropy
bifurcation. Obviousely, nonextensive systems are subject to a
dual nature since positive $\kappa$-values imply the tendency to
less organized states where the entropy increases whereas negativ
$\kappa$-values provide a higher organized state of decreased
entropy, see Leubner (\cite{Leubner05}).

Next, consider a DM halo as a self-gravitating collisionless system of
particles in dynamical equilibrium (e.g. Burkert \cite{Burkert00};
Firmani et al. \cite{Firmani00}). Consistent with Eq. (\ref{1})
the corresponding generalised entropy
$S(\kappa)$, characterising systems subject to long-range
interactions and couplings in nonextensive statistics, reads
(Tsallis \cite{Tsallis88}; Leubner \cite{Leubner04})

\begin{equation}
S_{\kappa }=\kappa k_{B}({\sum }p_{i}^{1-1/\kappa }-1)
\label{2}
\end{equation}

where $p_{i}$ is the probability of the $i^{th}$ microstate,
$k_{B}$ is Boltzmann's constant. The transformation $\kappa =
1/(1-q)$ links the $\kappa$-formalism, commonly applied in
astrophysical plasma modelling to the Tsallis q-statistics
(Leubner \cite{Leubner02}). Here $\kappa = \infty$ represents the
extensive limit of statistical independence and recovers the
classical BGS entropy as $S_{B}=-k_{B}{\sum p_{i}}\ln p_{i}$.

Since entropy and probability distributions reside physically on
the same level the corresponding generalized energy distribution,
accounting for long-range interactions, is available. In Maxwells
derivation the velocity components of the distribution $f(\bf{v})$
are uncorrelated where $ln f$ can be expressed as a sum of the
logarithms of the one dimensional distribution functions. In
nonextensive systems one needs to keep correlations between the
components, which can be done conveniently by extremizing the
entropy under conservation of mass and energy yielding the
corresponding one- and three-dimensional power-law distributions
in velocity space (Leubner \cite{Leubner04})

Here we must retain the spatial dependence and consider a
spherical symmetric, self-gravitating and collisionless N-body
system where the corresponding steady state phase-space
distribution $f(r,v)$ obeys the Vlasov equation. If the system of
particles itself provides the gravitational potential $\Phi$ and
$f(r,v)$ is regarded as the mass distribution then Poisson's
equation reads

\begin{equation}
\Delta \Phi = 4\pi G\rho = 4\pi G\int f(\frac{1}{2}v^{2} +\Phi
)d^{3}v, \label{3}
\end{equation}

This representation governs the equilibrium of the system where
commonly the relative particle energy $E_{r} = -1/2v^2 + \Psi$ is
introduced (e.g. Binney \& Tremaine \cite{Binney94}).

Extremizing the generalised entropy functional (\ref{2}) with
regard to conservation of mass and energy (Plastino
\cite{Plastino93}) the resulting energy distribution
$f^{\pm}(E_{r})$ involved in Eq. (\ref{3}) reads

\begin{equation}
f^{\pm}(E_{r})=B^{\pm} \left[ 1+\frac{1}{\kappa}
\frac{v^{2}/2-\Psi} {\sigma^{2}}\right] ^{-\kappa}
\label{4}
\end{equation}

The superscripts refer to the positive or negative intervals of
the entropic index $\kappa$, accounting for less (+) and higher
(-) organized states and thus reflecting the accompanying entropy
increase or decrease, respectively, (see Leubner
\cite{Leubner05}). $\sigma$ represents the mean energy or
variance of the distribution and $B^{\pm}$ denote the
corresponding normalisation constants depending on the entropic
index $\kappa$, for details see Leubner (\cite{Leubner04}). For
$\kappa \rightarrow \infty$ Eq. (\ref{4}) approaches the
exponential distribution function defining the density profile of
the isothermal sphere (Binney \& Tremaine \cite{Binney94}).

After incorporating the sign of $\kappa$ into Eq. (\ref{4}), we
perform separately for positive and negative definite $\kappa$ the
integration over all velocities where $B^{\pm}$ must be used
consistently. The resulting solution provides the relation for the
density evolution of a system subject to long range interactions
in a gravitational potential as

\begin{equation}
\rho^{\pm}=\rho_{0}  \left[ 1-\frac{1}{\kappa} \frac{\Psi} {\sigma^{2}}\right] ^{3/2-\kappa}
\label{5}
\end{equation}

Eq. (\ref{5}) generates for finite
positive values of $\kappa$ pronounced density tails, whereas for
negative $\kappa$-values the solutions are restricted within the
cutoff at $\kappa = \Psi / \sigma^{2}$ and $\kappa = -\infty$.

The duality of equilibria in nonextensive statistics is manifest
in two families, the nonextensive thermodynamic equilibria and the
equilibria of kinetic equations, where positive $\kappa$-values
correspond to the stationary states of thermodynamics and negative
$\kappa$-values to kinetic stationary states (Karlin et al.
\cite{Karlin02}). The limiting BGS state for $\kappa = \infty$
is characterized by self-duality. Physically the parameter
$\kappa$ is related to the heat capacity of the medium where
negative heat capacity, corresponding to values $\kappa < 0$,
is a typical property of self-gravitating systems, see e.g.
Firmani et al. (\cite{Firmani00}). The nonextensive bifurcation
into two distributions $f^{\pm}(E_{r})$ or $\rho^{\pm}$, respectively,
requires to identify the  density profile (\ref{5}) for positive
definite $\kappa$ as the proper distribution of the thermodynamic
state of the gas, whereas the negative definite counterpart is
associated with the self-gravitating DM distribution, we will
focus on. In the limit $\kappa \rightarrow \infty$ both solutions
merge at the isothermal sphere solution, defined by the limiting
exponential distribution obtained for $\kappa = \infty$
in Eq. (\ref{4}).

Finally, we combine Poisson's equation $\Delta \Psi = -4\pi G\rho$
and the density distribution Eq. (\ref{5}), which yields after
re-arranging for $\Psi$ a second order nonlinear differential
equation for the variation of a spherically symmetric gas and DM
distribution as (Leubner \cite{Leubner05})

\begin{equation}
\frac{d^{2} \rho}{{dr}^{2}}+\frac{2}{r} \frac{d\rho}{dr}-(1-\frac{1}{n})
\frac{1}{\rho}(\frac{d\rho}{dr})^{2}- \frac{4\pi Gn}{(3/2 - n) \sigma^{2}}
\rho^{2}(\frac{\rho}{\rho_{0}})^{-1/n}=0
\label{6}
\end{equation}

where $n=3/2-\kappa$ is introduced and corresponds to the
polytropic index of stellar dynamical systems. For negative
definite $\kappa$-values the DM density profiles defined by Eq.
(\ref{6}) are found numerically by solving the corresponding
two first order differential equations with a Runge-Kutta routine.
\begin{figure}
\begin{center}
{\includegraphics[width=\columnwidth]{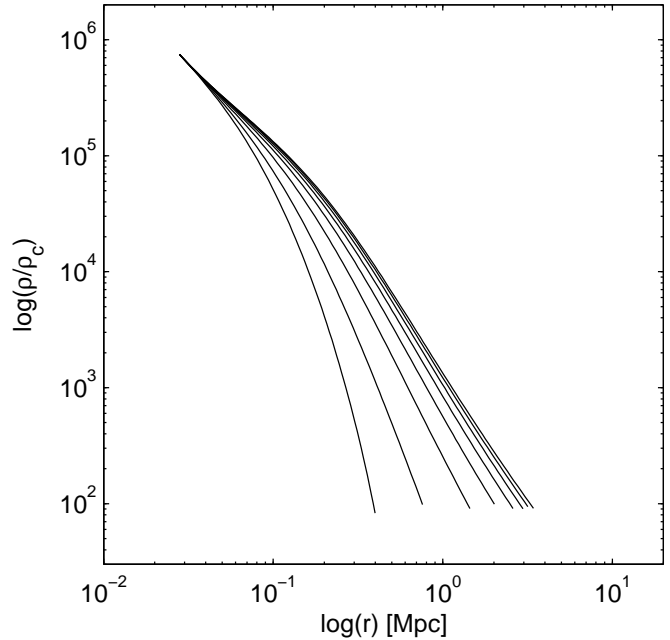}}
\caption{Characteristics of the radial DM density profile: A
sequence of solutions to Eq. (\ref{6}) is plotted corresponding to
different values of the parameter $\kappa$, varied from -2 to
$-\infty$ (left to right). The second free parameter $\sigma$ was
kept fixed for this schematic plot.} \label{fig1}
\end{center}
\end{figure}
Fig. \ref{fig1} shows the characteristics of the radial density profile
as depending on the variation of the entropic index $\kappa$ and computed from Eq.
(\ref{6}) for DM profiles. With increasing $\kappa$ the solution
converges from left to right to the isothermal sphere density profile,
an environment subject to statistical independence of the constituents.

\section{N-body simulations}

The numerical simulations were performed using an adapted version
of the treecode of Barnes \& Hut (\cite{Barnes86}), using 128$^3$
particles in a comoving volume of (32 h$^{-1}$ Mpc)$^3$ in a
$\Lambda$CDM cosmology with $\Omega_\Lambda=0.7$, $\Omega_{\rm
m}=0.3, h=0.7, \sigma_8=0.93$. We employ the constrained random
field technique of Hoffman \& Ribak (\cite{Hoffman91}), as
implemented by van de Weygaert \& Bertschinger
(\cite{Weygaert96}), to form a rich cluster in the centre of the
simulation box. The (Plummer) softening length was set to 7
h$^{-1}$ kpc, which also sets the spatial resolution of the
simulation. Well over a thousand time steps were used to integrate
the constrained initial conditions to the present epoch. The
diagonal components of the velocity dispersion tensor for the mass
within 3 Mpc are: $\sigma_x = 804$ km/s, $\sigma_y = 872$ km/s,
and $\sigma_z = 1043$ km/s.

To obtain a spherically averaged mass density profile at each time
step, we first determine the cluster centre using an iterative
technique similar to the one presented in Power et al.
(\cite{Power03}). We compute the centre of mass from particles
within spheres of decreasing radius. Then we bin the particles
logarithmically according to their distance from the cluster
centre and calculate the density profile out to a radius of $\sim
2.5$ Mpc (physical distance). In order to get reliable results, we
consider only scales larger than the softening length of the
particles and the number of particles within each bin is always
well above 100.
\begin{figure}
\begin{center}
{\includegraphics[width=\columnwidth]{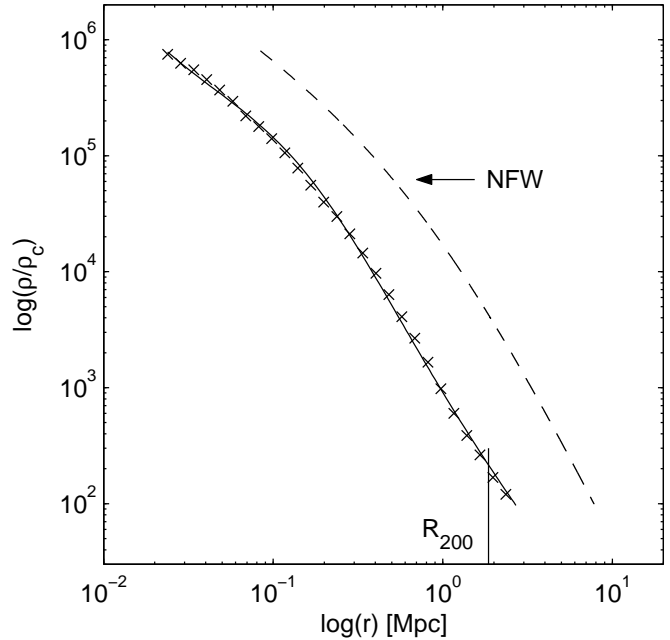}}
\caption{Radial dark matter density profile obtained from the
simulation (crosses). The solid line shows a fit of the
nonextensive theory to the data with best fitting values of
$\kappa = -15$ and $\sigma=0.12$. For comparison, also the best
fitting NFW profile is provided as a dashed line but shifted to
the right for better visibility. $R_{200}$ indicates the virial
radius.} \label{fig2}
\end{center}
\end{figure}

\section{Discussion and conclusions}

With this setup we studied a sample of 3 cluster-sized DM haloes
with different formation histories. We confirm the commonly known
property that the inner density profile of the halo does not
evolve anymore after a redshift of about one. We then use the
relaxed density profile of the most massive cluster halo of our
sample (M$_{vir}$=8.2$\times10^{14}$ M$_{\odot}$ within R$_{200}$
at z=0), which is also the most virialised system already at
z$\approx$1, and fit the nonextensive distribution by solving the
differential equation (\ref{6}). The result is shown in Fig.
\ref{fig2} together with the corresponding NFW profile for
comparison. The nonextensive density profile follows the result of
the simulation accurately at all cluster radii with best fitting
values of $\kappa = -15$, indicating the presence of strong
correlations in the system, and $\sigma = 0.12$. Hansen et al.
(\cite{Hansen05}) investigated possible theoretical constraints on
the entropic index $q$, or equivalently $\kappa$. Arguing that
negative velocity dispersions are difficult to interpret in a
cosmological context, he arrived at the condition $q<5/3$, which
reads in the $\kappa$ formalism $\kappa<-3/2$. By demanding a
positive proportion between temperature and internal energy
Boghosian (1999) arrived at the same constraint. Our best fitting
value of $\kappa = -15$ clearly obeys this restriction.

\begin{figure}
\begin{center}
{\includegraphics[width=\columnwidth]{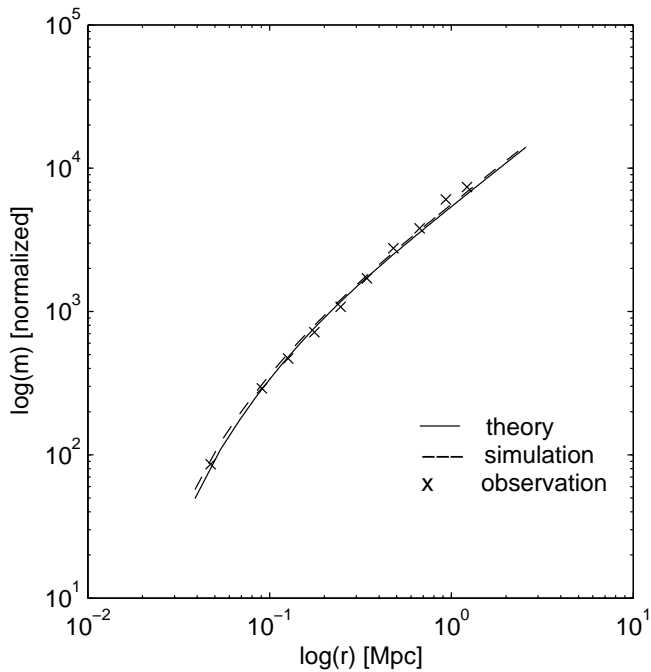}}
\caption{Integrated mass profile of the galaxy cluster A1413
(crosses, Pointecouteau et al. \cite{Pointecouteau05}) compared to
the simulated (dashed) and theoretical (solid) curve.}
\label{fig3}
\end{center}
\end{figure}

Fig. \ref{fig2} demonstrates that also the empirical NFW profile
represents the simulated radial density distribution well.
However, the use of the nonextensive approach allows one to
interpret differences that might occur, e.g. in comparison to
observations, within the physical context of entropy
generalisation. On the other hand, in deriving Eq. (\ref{6}) we
assumed that the phase-space distribution function $f(r,v)$ is a
function of energy only. This is essentially true for systems
having an isotropic velocity dispersion tensor. If this is not the
case, as suggested by numerous numerical simulations (e.g. Cole \&
Lacey \cite{Cole96}; Natarajan, Hjorth \& van Kampen
\cite{Natarajan97}), $f(r,v)$ is generally also a function of the
angular momentum vector $\bf{L}$ of the system. Therefore, for
systems with a highly anisotropic velocity dispersion, solutions
of Eq. (\ref{6}) might not represent the density profile in
detail. Furthermore, there is no physical reason to keep $\kappa$
constant, as used within the present analysis. For instance, one
might suspect $\kappa$ to vary with radius, as the strength of the
correlations could be a function of the distance. However, since
the shape of the density profiles, from simulations and
observations, are reproduced accurately within the applied
simplifications (see Figs. \ref{fig2} and \ref{fig3}), and given
the fact that also the cluster halo shown in Fig. \ref{fig2} is
not completely isotropic (see section 4), modifications due to
these open issues are expected to be small. In addition, presently
the resolution of the simulation is insufficient to obtain robust
estimates for the density distribution at smaller radii than
presented in Fig. \ref{fig2}. However, from high resolution
simulations (e.g. Navarro et al. \cite{Navarro04}) it is known
that DM profiles do not become flat at currently probed radii
($<$1\% of R$_{200}$).
Finally, in Fig. \ref{fig3} the present analysis is tested on an
integrated mass profile evaluated from X-ray observations. Along
with the integrated simulation profile and the nonextensive
solution a normalized profile of the galaxy cluster A1413 is
plotted (Pratt \& Arnaud \cite{Pratt02}; Pointecouteau et al.
\cite{Pointecouteau05}). In the observable radial range both, the
simulated and the nonextensive profile, perfectly match the
observed data set. Note that A1413 is a relaxed cluster with no
cooling core, consistent with the theoretical context restricted
to equilibrium states.

In principle, as the topology of the solutions to equation
(\ref{6}) is very complex, also density profiles from systems in
earlier evolutionary stages can be well reproduced. However, in
deriving the differential equation (\ref{6}) we implied that the
system is relaxed, as we extremized the generalised entropy
functional (\ref{2}) to get the equilibrium phase space
distribution function. As the cluster is not relaxed at redshifts
$z>1$ the physical interpretation of the fitting parameters fails
or is extremely complicated in this redshift range.

We also note that the strength of correlations within a system
depends on the cosmological model, or more precisely on the form
of the power spectrum of the initial density fluctuations, i.e.
$P(k) \propto k^n$. It has been recognized already in early
numerical works that increasing $n$ leads to a steepening of the
density profiles (e.g. Hoffman \cite{Hoffman88}; Efstathiou et al.
\cite{Efstathiou88}). Our theoretical framework of non-extensive
statistics and the derivation of Eq. (\ref{6}) do not depend on
the initial power spectrum. Therefore the theory is also valid for
different power spectra than the one adopted for this work
($n=1$). Even without performing additional simulations ourselves,
we can conclude that long-range interactions are more important
for high spectral indices $n$, as steeper profiles are realised by
lower values of $\kappa$.

In conclusion, nonextensive statistics provides on physical
grounds access to the study of DM density profiles in relaxed
clusters and is able to model the corresponding equilibrium states
of self gravitating collisionless systems. The significant
advantage over empirical fitting functions is provided through the
physical content of the parameters involved in the nonextensive
approach, with $\kappa$ as a measure of the degree of correlations
in the system and $\sigma$, the characteristic energy. Links
between the entropic index $\kappa$, the heat capacity and the
polytropic index of DM haloes are subject of a further study.

\section*{Acknowledgements}
The authors acknowledge the German Science Foundation (DFG),
through grant Zi 663/6-1, the Austrian Science Foundation (FWF),
through grant P15868 and the UniInfrastrukturprogramm 2004 des
bm:bwk Forschungsprojekt Konsortium Hochleistungsrechnen. We are
grateful to the anonymous referee for his/her criticism that
helped to improve the paper. Joshua Barnes and Piet Hut are
gratefully acknowledged for use of their treecode, and Edmund
Bertschinger and Rien van de Weygaert for use of their constrained
initial conditions code. Etienne Pointecouteau, Gabriel Pratt and
Monique Arnaud are gratefully acknowledged for providing the data
of the integrated mass profile of the cluster A1413.

\end{document}